%% file: main.tex
\documentclass[aic]{iosart2x}
\numberlinesfalse
\usepackage{graphicx}
\usepackage{algorithmic}
\usepackage{algorithm}
\usepackage{textcomp}
\usepackage{booktabs}
\usepackage{multirow}
\usepackage{diagbox}
\usepackage{color}
\usepackage{colortbl}
\usepackage{subfigure}
\usepackage{xcolor}
\usepackage{xltabular}
\usepackage{multicol}
\usepackage{listings}
\usepackage{stfloats}
\usepackage{lipsum}

\input{bnf_macros}

\usepackage{caption}
\usepackage{subcaption}
\usepackage{subfigure}
\usepackage{amsmath}
\usepackage{enumitem}
\pubyear{0000}
\volume{0}
\firstpage{1}
\lastpage{1}

\begin{document}
\begin{frontmatter}
%\pretitle{}
\title{\centering A methodology and a platform for high-quality rich personal data collection\thanks{This paper is a major revision and extension of \cite{kayongo2024methodology}.}}
% \title{Enhancing Human Awareness Through Big-Thick Data Collection}

%Human Awareness Model to achieve Big-Thick Data Collection

%Human-in-the-loop Big-Thick Data Collection

%\runtitle{Context-Aware Scheduler System for Mobile Self-Reports}
%\subtitle{}

% For one author:
%\author{\inits{N.}\fnms{Name1} \snm{Surname1}\ead[label=e1]{first@somewhere.com}}
%\address{Department first, \orgname{University or Company name},
%Abbreviate US states, \cny{Country}\printead[presep={\\}]{e1}}

% Two or more authors:
\begin{aug}
\author{\inits{I.}\fnms{Ivan} \snm{Kayongo}\ead[label=e1]{ivan.kayongo@unitn.it}%
%\thanks{Corresponding author. \printead{e1}.}
}
\author{\inits{N.}\fnms{Leonardo} \snm{Malcotti}\ead[label=e2]{leonardo.malcotti@studenti.unitn.it}}
\author{\inits{N.}\fnms{Haonan} \snm{Zhao}\ead[label=e3]{haonan.zhao@unitn.it}}
\author{\inits{N.}\fnms{Fausto} \snm{Giunchiglia}\ead[label=e4]{fausto.giunchiglia@unitn.it}}
\address{Department of Information Engineering and Computer Science, \orgname{University of Trento}, \cny{Italy}\printead[presep={\\}]{e1,e2,e3,e4}}
%\address[B]{Department first, \orgname{University or Company name},
%Abbreviate US states, \cny{Country}\printead[presep={\\}]{e2,e3}}
\end{aug}

\begin{abstract}
In the last years the pervasive use of sensors, as they exist in smart devices, e.g., phones, watches, medical devices, has increased dramatically the availability of personal data. However, existing research on data collection primarily focuses on the objective view of reality, as provided, for instance, by sensors, often neglecting the integration of subjective human input, as provided, for instance, by user answers to questionnaires. This limits substantially the exploitability of the collected data. In this paper we present a methodology and a platform specifically designed for the collection of a combination of large-scale sensor data  and qualitative human feedback. The methodology has been designed to be deployed on top, and enriches the functionalities of, an existing data collection APP, called \textit{iLog}, which has been used in large scale, worldwide data collection experiments. The  main goal is to put the key actors involved in an experiment, i.e., the \textit{researcher} in charge, the \textit{participant}, and \textit{iLog} in better control of the experiment itself, thus enabling a much improved quality and richness of the data collected. The novel functionalities of the resulting platform are:  (i) a time-wise representation of the situational context within which the data collection is performed, (ii) an explicit representation of the temporal context within which the data collection is performed, (iii) a calendar-based dashboard for the real-time monitoring of the data collection context(s), and, finally, (iv) a mechanism for the run-time revision of the data collection plan.  The practicality and utility of the proposed functionalities are demonstrated by showing how they apply to a case study involving 350 University students.

\end{abstract}


% \begin{keyword}
% \kwd{Human-Centered Studies}
% \kwd{Human awareness Model}
% \kwd{ESM}
% \kwd{EMA}
% \kwd{User Experience}
% \kwd{Logical Architecture}
% \end{keyword}

\end{frontmatter}

%%%%%%%%%%% The article body starts:

% \input{section/1.Introduction}
\input{section/1-Introduction}

\input{section/2-Related_Work}

\input{section/2.5-Motivating-example}

\input{section/3-Representing_context}

\input{section/4-Scheduling_plan}

\input{section/5-Monitoring_plan_execution}

\input{section/6-Learning}

\input{section/8-Discussion}

 \section*{Acknowledgment}

The research by Fausto, Ivan, and Leonardo were funded by the European Union's Horizon 2020 FET Proactive project “WeNet – The Internet of us” (\texttt{\url{https://internetofus.eu/}}), grant agreement No 823783. 
The work by Haonan received funding from the China Scholarships Council (No.202107820038). This work has been possible only thanks to the interactions and feedback from the people working in WeNet. 

% ---- Bibliography ----

% if your bibliography is in bibtex format, use those commands:
\bibliographystyle{ios1}       
\bibliography{main}    

\end{document}

%% file: bnf_macros.tex
\usepackage{xspace}
\newcommand{\bnfdef}{\texttt{:= }}

\newcommand{\bnfterm}[1]{\texttt{<#1>}}
\newcommand{\bnfor}{$|$ }

\newcommand{\bnfend}{;\\}

%% file: section/1-Introduction.tex
\section{Introduction}
% \noindent
In today’s world, digital interactions have become deeply integrated into daily life, generating vast amounts of personal data. This data, encompassing information about individual identities, preferences, activities, and interactions, is increasingly collected through digital devices, online services, and various monitoring technologies. For instance, smart devices, e.g., phones, watches, or medical devices, are equipped with numerous sensors that collect massive volumes of data about their owners. This type of data, often referred to as \textit{big data} \cite{das2013big}, is characterized by its vast volume, high velocity, and diverse variety, and allows for the identification of large-scale patterns and trends through advanced computational techniques. Despite its power, big data often lacks the contextual depth needed to fully understand the underlying human elements behind the numbers, that is, it fails to explain the
subjective impulses that drive an individual’s actions. 
In contrast to the vast amount and speed of "big data",  \textit{thick data} provides a qualitative description focused on human experience and behavior \cite{geertz2008thick}. Thick data pertains to the abundant and detailed insights obtained from extensive qualitative research techniques such as ethnography, interviews, and participant observation. It prioritizes qualitative aspects such as human narratives, emotions, and cultural subtleties, i.e., it is a class of data sources that align with ethnographically collected and meticulously analyzed observational data.
Building upon the two notions above, \cite{bornakke2018big} defines \textit{big-thick data} as the convergence of \textit{big-thin data}, e.g., usage analytics, sensor data, general Internet-of-Things (IoT) data, with \textit{small-thick data}, e.g., observations, interviews and questionnaires. The intuition is that big quantitative data, prized for its objectivity and scalability, complements the contextual richness of the qualitative insights of thick data \cite{d2023data}. 
 
 The notion of big-thick data was originally conceived having in mind the human-centric design of services, with the goal of blending statistical rigor with contextual relevance \cite{bogers2016connected, bogers2018situated}.
 However, this intuition is very powerful and can be applied in almost all AI applications, and Machine Learning (ML) in particular. In ML, for instance, the user's subjective interpretation of the current situation can help the machine in building a better understanding of what is going on, for instance in order to enable better human-machine interactions \cite{KD-2022-Bontempelli-lifelong} or better machine-enabled social interactions \cite{OsmanCSSG21,KD-2021-WeNetDiversityOne}. The integration of these two types of data facilitates the meaningful bi-directional human-machine collaboration by providing data that allows the machine to learn from human behavior and activities, as well as data that captures the human interpretation of their actions. It is not by chance that \cite{bornakke2018big} mentions  social media and Experience-Sampling-Method (ESM) data as early examples of big-thick data.
 Building upon this intuition, \cite{giunchiglia2024big} shows how big-thick data can be generated by integrating context-aware personal data, collected using both sensors and questionnaires, with data about the environment within which the personal data are collected, this being done by exploiting OpenStreetMap (OSM) data enriched with other datasets carrying detailed information about the places involved. As shown in  \cite{giunchiglia2024big}, the result is a very rich dataset which, while being more focused and much smaller than the original datasets, allows to learn about and provide answers to a much richer set of questions which integrate the objective view with one or more personal subjective views of the current situation.
 
However, at the current state of the art, the \textit{quality of the data collected from users} is a major limitation which limits the generation of big-thick data, when this is not done manually but, rather, delegated to a data collection APP.
The goal of this paper is \textit{to describe a methodology and a platform that enables participants to provide high-quality personal data,} as close as possible to the richness of big-thick data, \textit{while ensuring minimal disturbance to the user.} The target are all the researchers who have a need for the kind of data we want to produce. At the moment we have identified at least four such groups: (i) Researchers in AI and ML with a focus on personalized services; (ii) Computational Social Science researchers where the subjective component is key; (iii) Psychology researchers and in particular those following the EMA /ESM methodology (see the related work section) and (iv) service design researchers with a focusing on the problem of \textit{designing with data}. See, for instance the hackathon described in \footnote{\url{https://datascientiafoundation.github.io/diversityone-2025/}.}

The starting point of the methodology is the identification of the three key roles around which the data collection process evolves, that is: (i) the \textit{researcher}, that is the person who has designed the experiment and that, during its execution, monitors its evolution, (ii) the \textit{participant} to the experiment, one or more, where also the researcher can be a participant, namely the person in charge of providing data, via one or more mobile devices, the data to be collected, and, finally, (iii) the \textit{platform}, collecting the data from the participants. The  intuition is to develop a set of features, and corresponding mechanisms, where these three roles have increased awareness and control over the data collection process. 
We instantiate this intuition via four key functionalities, each building upon and extending the previous one.
 
\begin{itemize}

\item \textit{A representation of the situational context} within which the data collection is performed.
The understanding of the local context (including the user's physical and psychological context) is key to the idea of thick data
\cite{geertz2008thick} and its relevance has been pointed out in most mobile applications, see, e.g., \cite{intille2003context,runyan2013smartphone,wang2014studentlife,huang2016assessing,KD-2021-Zhang-putting}. 
Knowledge of the user’s physical, social, emotional, and informational states, allows to better interpret the vast amounts of sensor data collected \cite{chen2012business}, thus improving the relevance and quality of data collected \cite{davenport2013big,boyd2012critical}. The key innovation in this paper is that we focus on the context of the data collection as such. The machine works in some kind of meta-context whose sole goal is to increase the machine / participant / researcher's awareness of the process of data collection, as a first step towards increasing control and the quality and richness of the collected data. That is, our ultimate aim  is \textit{to generate big-thick data about the process of generation of big-thick data}, the former being a key enabler for the generation of the latter.

\item \textit{A representation of the temporal context} within which the data collection is performed. By this we mean
that an experiment is modeled as a plan where each action, e.g., a human answer to a machine question, or a sensor data collection, or a machine answer to a human question, is associated with a set of \textit{scheduling constraints} and, after execution, with a set of \textit{execution annotations}, encoding information about past, present and future actions. Examples of planning constraints are, for instance, that a question can be asked within a certain time frame, and that should be asked only when at home. Examples of plan execution annotations are, for instance, that a question was not answered, of that it was answered with a delay of half an hour. To this extent,
we have developed a representation language, called \textit{iLogCal}, which allows to represent all the context dimensions, both temporal and situational, and to use them to condition the activation of both questions to the user and sensor data collection activation / stop. 

\item \textit{A calendar-based dashboard} which allows all the three roles to focus on specific elements 
of the experiment temporal and situational context. One of the key aspects is that iLogCal has been defined by extending (a subset of) \textit{iCal}, 
 the Internet standard Calendar\footnote{\url{https://www.ietf.org/rfc/rfc5545.txt}.}$^{,}$\footnote{\url{https://icalendar.org/}.}. This allows to use calendars to provide multiple views of a plan, for instance, by focusing on present, past or future, or on a specific context, or on a set of quality parameters, or on one or more participants, while maintaining an overall holistic view of the experiment.

\item \textit{A set of mechanisms for the execution-time revision of the data collection plan.} The data collection plan can be revised by a single participant, within the bounds  set by the researcher, or by the researcher for one or more of the participants. The plan can also be revised by the platform itself, for instance, based on a ML algorithm which has learned what are the best / worst dates for getting the answer of the best quality. 
%In case of contradictory suggestions, the researcher can fix a variation period (possibly null) within which the researcher can modify the plan while the decision of the machine can always be overwritten by the participant and, therefore, the researcher. 
The control hierarchy proceeds from the researcher, to the participants, to the platform.
\end{itemize}
These four functionalities are being implemented as part of an integrated platform, an APP, called \textit{iLog}, built on top of an earlier version  of \textit{iLog} itself \cite{zeni2014multi}. The two key core features of this earlier version of iLog are the possibility (i) of collecting sensor data from any number of sensors from one of more smart devices, and (ii) of collecting user-provided answers to questionnaires, which can be synchronic as well as diachronic. Since its first application in 2013, iLog has been used in many data collections campaigns, see, e.g., \cite{KD-2019-Maddalena-WWW,bison2021trento,KD-2020-Zeni1,giunchiglia2021worldwide}. These experiments have allowed to generate an extensive set of studies, see e.g., \cite{bontempelli2021learning,assi2023complex,meegahapola2023generalization,giunchiglia2024big}, while at the same time highlighting the problems of data quality that motivate this work. This paper is a rather detailed description of the four functionalities described above and of how they are integrated, as part of a single platform, on top of the original version of iLog.

This paper is structured as follows. Section \ref{relwork} decribes the related work. Section \ref{study} introduces the main features of iLog and how it was used in an experiment, described in \cite{giunchiglia2021worldwide}, carried out as part of the WeNet project \cite{WenetPaper}\footnote{\url{https://www.internetofus.eu/}.}.  The description is rather concise focusing only on those aspects which relate to the four functionalities above.
A detailed description of the resulting dataset (including GDPR and ethics compliance)\footnote{The dataset can be downloaded from the DataScientia Web site \url{https://datascientia.disi.unitn.it/}.}, is provided in \cite{bussoDiversityOne}.
 Section \ref{repctx} describes the situational context model. Section \ref{scheduling1} introduces the main features of the temporal context model and iLogCal. Section \ref{monitoring} focuses on the monitoring process. Section \ref{learning} provides an example where the ML component improves the answer quality based on an analysis of how the temporal and situational context (and a few other parameters) influence how long a participant waits before starting to answer a question. 
 %Section \ref{arc} provides a high level view of how iLoG is extended to produce a platform aggregating the four modules implementing the functionalities described above. 
 The notions from Sections \ref{repctx}, \ref{scheduling1}, \ref{monitoring}, \ref{learning}  are exemplified on the experiment and dataset described in Section \ref{study}. Finally, Section \ref{discussion} closes the paper.

%% file: section/2-Related_Work.tex
\section{Related Work}
\label{relwork}

The intuition underlying the notion of context is very similar to that underlying the notion of big-thick data. That is, the knowledge of the local situation is key in order to provide machines with a good enough understanding of what is going on. This notion has been extensively studied and most early studies on context were in Knowledge Representation (KR) and AI, \cite{mccarthy1987generality,KD-1993-giunchiglia}. Later on, Schilit et al. \cite{schilit1994disseminating} introduced the concept of context, defining it as involving “locations, identities of nearby individuals and objects, and changes to those objects.” Similarly, Brown et al. \cite{brown1997context} depicted context as being about “locations, varying user roles, time, seasons.” In \cite{dey1998cyberdesk}, Dey et al. provides a definition of context which is closely aligned with our understanding as, “the user’s physical, social, emotional, or informational states.” Dey and Abowd \cite{abowd1999towards} define the context in a more comprehensive manner. They state that “context is any information that can be used to characterize the situation of an entity. An entity is a person or object that is considered relevant to the interaction between a user and an application, including the user and applications themselves.” 
 Existing research on the development of context-aware applications primarily concentrates on interruptions in context \cite{H-2017-Mishra}, gathering user attention \cite{mehrotra2016my}, or enhancing the response rate of questions \cite{H-2021-Sun}. These studies underscore the necessity of providing users with suitable times to facilitate user acceptance of information, with instances found in expert systems \cite{ye1995impact} and, more recently, reccommender systems \cite{afolabi2019improving}. 
 %This is because questions often arrive at inconvenient times, such as during meetings or while driving, disrupting daily routines and resulting in low-quality responses which in-turn affects the quality of the data collected. 
 %
 However, so far, no work has ever concentrated on the \textit{meta-context} of the process of data collection. In general, overall, there has been minimal focus on providing the user with flexibility in answering questions or in driving the sensor collection flow. Typically, the researcher's dispatch of questions consists of a fixed schedule, with no possibility for the researcher, or participant, or the platform, to control the data collection and in particular the participant's response activity. 
 
 Numerous systems, which leverage data from mobile devices and wearable sensors, have found applications in health monitoring, aging care \cite{lee2015sensor,berke2011objective}, and the understanding of human behaviors and traits. For instance, mobile sensing has proven invaluable for health and physical activity monitoring, where accelerometers, gyroscopes, and GPS sensors are used to track users’ movements and generate insights into exercise routines, sedentary behaviors, and overall activity levels \cite{dunton2012momentary,liao2015using,rabbi2015mybehavior,intille2016precision} and in research on comprehending and forecasting human behaviors and traits \cite{do2012contextual,farrahi2011discovering, harari2016using,wang2018sensing,peltonen2020phones}. Similarly, behavioral data such as sleep patterns, social interactions, and phone usage have been utilized to detect early signs of mental health issues like stress, depression, and anxiety \cite{bogomolov2014daily, wang2018tracking,wang2020social}. 
Some of this work has resulted in the development of several
mobile sensing frameworks designed to support the data collection and analysis.
% data collecting platforms, which are most pertinent to our work 
Some such examples are 
Campaignr \cite{joki2007campaignr} and Epicollect 5 \cite{gohil2020epicollect} which focus on customizable and scalable data collection; CenceMe \cite{miluzzo2008sensing} and Social fMRI \cite{aharony2011social}  which emphasize the social context sensing; Empath2 \cite{dickerson2015empath2}, Emotion Sense \cite{lathia2013smartphones}, and ESMAC \cite{bachmann2015esmac} which specialize in emotion detection and behavior analysis.

  Closer to our work, is AWARE \cite{ferreira2015aware}, a well-established mobile sensing framework designed for the collection of passive data through smartphone sensors. While AWARE provides a broad and extensible framework for environmental and behavioral context awareness, iLog excels in comprehensive activity tracking, making it a more effective tool for personalized research contexts where user-specific logging are critical.
  DemaWare2 \cite{stavropoulos2017demaware2} is another prominent framework designed for activity recognition and contextual reasoning. It uses sensor fusion and ontologies for the detection of complex activities. While DemaWare2 excels in identifying hierarchical activities through predefined rules and ontologies, iLog offers greater flexibility.  DemaWare2's focus on predefined activity hierarchies may not be as effective in capturing more fluid or personalized data, thus limiting its applicability in dynamic, real-world settings.
  The Effortless Assessment Research System (EARS) \cite{lind2023reintroducing} is designed for passive monitoring of behavioral and psychological patterns, particularly in mental health research. It emphasizes effortless data collection by minimizing the need for active user participation. While this makes EARS ideal for studies requiring low participant burden, it may miss opportunities for richer contextual insights that can be derived from active input. 
  Beiwe \cite{beiwe} is a high-throughput digital phenotyping platform designed for mental health research and behavioral studies. Like iLog, it integrates passive sensing with active data collection (through surveys), allowing for personalized data collection and analysis. However, iLog offers a broader scope of application, extending beyond the mental health focus of Beiwe to include areas such as habit formation and lifestyle monitoring. Moreover, iLog stands out due to its emphasis on user engagement and ethical data collection practices, providing greater transparency and user autonomy over data sharing, which may not be as explicitly emphasized in Beiwe. 
  RADAR-base \cite{radar-base} is an open-source platform designed for longitudinal health studies, particularly in neurological and psychiatric research. It allows for the integration of multiple wearable sensors and mobile applications for the collection of health-related data. While RADAR-base excels in health-related contexts, iLog offers more flexibility across a broader spectrum of research domains. By combining passive sensor data with active surveys, iLog provides a richer, more personalized understanding of user behavior, mainly because of its ability to adjust data collection based on the context and the individual’s responses.

  % In summary, iLog offers several distinct advantages over other mobile sensing frameworks since 
  % % like AWARE, DemaWare2, EARS, Beiwe, and RADAR-base. 
  % its hybrid data collection method, combining both passive sensing and active user input, provides richer, more personalized insights into user behaviors. Additionally, iLog's flexibility in adapting to different research domains, its focus on ethical data collection, and its ability to offer dynamic, context-sensitive surveys make it a more user-centered and versatile platform compared to others in the field. As such, iLog stands out as a powerful tool for a wide range of research applications, from health monitoring to habit analysis and personalized services.

%% file: section/2.5-Motivating-example.tex
\begin{figure}
    \centering    \includegraphics[width=0.8\linewidth]{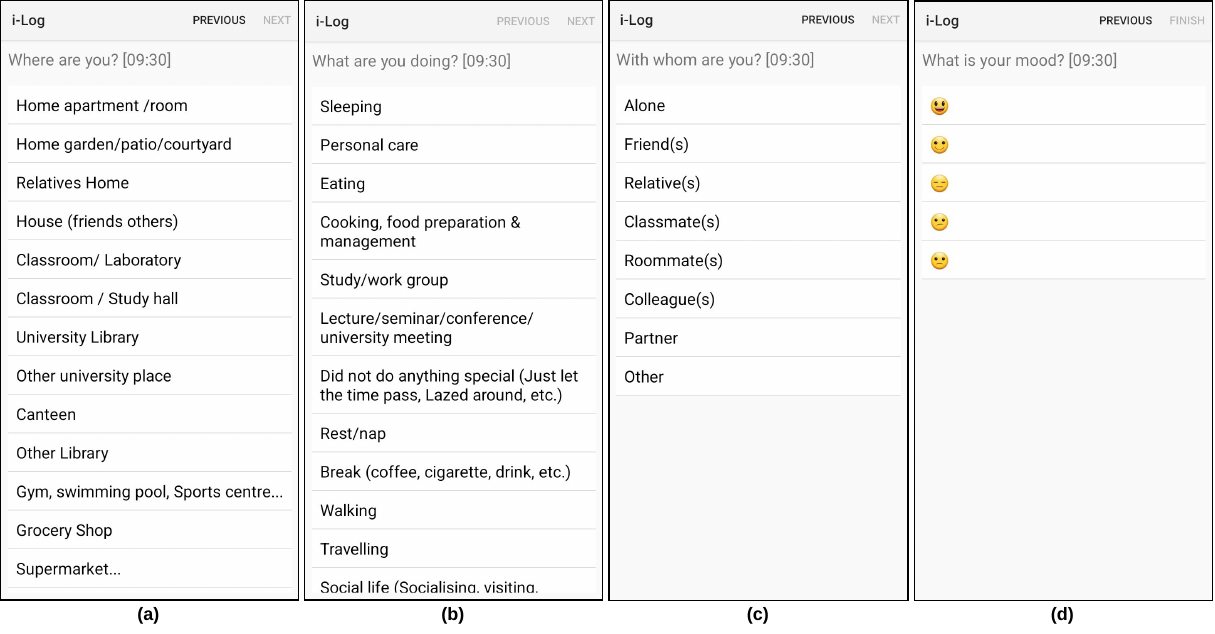}
    \caption{Sample questions captured in the WeNet project}
    \label{fig:sample_qnz}
\end{figure}

\section{iLog at work}
\label{study}

The experiment that generated the data that we consider here as our motivating example involved students from the University of Trento, Italy. The experiment was conceived and designed following the mainstream approach in the Social Sciences, in particular in the development of  \textit{time diaries}, where the questions submitted to participants follow the HETUS standard \footnote{More information about the \textit{HETUS (Harmonized European Time Use Surveys} standard can be found at \url{https://ec.europa.eu/eurostat/web/time-use-surveys.}}. \cite{KD-2020-Zeni1} describes a large scale European level data collection experiment which was used to fine tune the methodology used in this experiment. Following this methodology, as a first step, all students were invited via email to participate in a survey. From the 5,000+ respondents, a representative sample of 350 students was selected based on their fields of study and socio-demographic characteristics. To mitigate bias and noise, in this paper we consider the data of a selection of 170 students, where the choice is motivated by considerations related to the number of answers provided and demographic characteristics, including gender, study degree, and department (see Table \ref{socio}).
iLog runs on the participants' smart phones, both Android and iOS, and acts as an interface through which it is possible to capture annotations / tags / answers from participants.
  iLog allows for a wide range of question types (e.g., free text, fixed answers, take photo). The questions are sent at intervals defined inside the experiment plan. 
The  experiment described here consisted of three diachronic time-diaries,  with varying timings and aims:

\begin{table}[htpb!]
\centering
\caption{Selected participants' demographic information.}
\begin{tabular}{|c|cc|cc|cccc|}
\hline
  & \multicolumn{2}{c|}{\textbf{Sex}} & \multicolumn{2}{c|}{\textbf{Degree}} & \multicolumn{4}{c|}{\textbf{Department}}                 \\ \hline
Feature & Female          & Male            & BSc               & MA+PhD           & Information Science & Industrial & Business & Sociology \\
Number  &  101             & 69              & 108               & 62               & 50                  & 25         & 44       & 51        \\
 \hline
\end{tabular}
\label{socio}
\end{table}

\begin{enumerate}
    \item The first diary gathered general information about the day. At 08:00 AM, the participant received two qualitative questions about sleep quality and expectations for the day. At 10:00 PM, the participants were asked  (A) to rate their day; (B) to identify any problems they encountered during the day;  (C) to describe how they solved them; and, finally, (D) they received a question about the COVID-19 pandemic.

    \vspace{0.1cm}
    \item The second diary is a standard time diary with questions about three main activities and mood. Every half an hour for the first two weeks and every hour for the subsequent two weeks, the participants received a smartphone notification with four questions (the first three based on the HETUS standard):
    \begin{itemize}
        \item “What are you doing?” allowing for 34 different answers; 
        \item  “Where are you?” allowing for 26 different answers; 
        \item “Who is with you?” allowing for 8 different answers (including "being alone");  
        \item “What is your mood?” allowing for a scale of 5 levels ranging from happy to sad.
    \end{itemize}
    \item In the third time diary, the participants received an additional set of questions about food and drinks. These questions were asked every two hours outside the main meal hours.
\end{enumerate}
Fig. \ref{fig:sample_qnz} shows some of the questions (and pre-compiled answers) asked during the data collection.
iLog automatically collects sensor data in the background without requiring user intervention. Researchers are given the flexibility to design the frequency of data collection for each single sensor. In total, iLog allows to collect data from 34 sensors which are categorized  into three groups as follows:

\begin{itemize}
    \item Hardware (HW) sensors, the sensors typically found in smart phones, collect information about the surrounding environment.
     See  Table \ref{hw_sensors} for the list of HW sensors collected during the experiment.
\item Software (SW) sensors, also typically found in smart phones, collect data about the SW events involving the Operating system and the APPs. See Table \ref{sw_sensors} for list of SW sensors collected during the experiment.

    \item Question-Answering (QA) sensors collect information about the events that are associated with the question answering process. See Table \ref{questionnaire_sensors} for list of QA sensors.
\end{itemize}

%\vspace{-0.3cm}
\begin{table}[htb!]
\centering
\caption{HW sensors}
\begin{tabular}{|c|l|l|l|}
\hline
\textbf{No} & \textbf{HW Sensor} & \textbf{Estimated Frequency} & \textbf{Category} \\
\hline
1 & Accelerometer & up to 10 samples per second & Big \\
\hline
2 & Gyroscope & up to 10 samples per second & Big \\
\hline
3 & Light & up to 10 samples per second & Big \\
\hline
4 & Location & Once every minute & Small \\
\hline
5 & Magnetic Field & up to 10 samples per second & Big \\
\hline
6 & Pressure & up to 10 samples per second & Big \\
\hline
\end{tabular}
\label{hw_sensors}
\end{table}

 %\vspace{-0.1cm}
\begin{table}[htb!]
\centering
\caption{SW sensors}
\begin{tabular}{|c|l|l|l|}
\hline
\textbf{No} & \textbf{SW Sensor} & \textbf{Estimated Frequency} & \textbf{Category} \\
\hline
7  & Airplane Mode [ON/OFF] & On change & Small \\
\hline
8  & Battery Charge [ON/OFF] & On change & Small \\
\hline
9  & Battery Level & On change & Small \\
\hline
10 & Bluetooth Devices & Once every minute & Small \\
\hline
11 & Bluetooth LE (Low Energy) Devices & Once every minute & Small \\
\hline
12 & Cellular network info & Once every minute & Small \\
\hline
13 & Doze Mode [ON/OFF] & On change & Small \\
\hline
14 & Headset Status [ON/OFF] & On change & Small \\
\hline
15 & Movement Activity Label & Once every 30 seconds & Small \\
\hline
16 & Movement Activity per Time & Once every 30 seconds & Small \\
\hline
17 & Music Playback (no track information) & On change & Small \\
\hline
18 & Notifications received & On change & Small \\
\hline
19 & Proximity & up to 10 samples per second & Small \\
\hline
20 & Ring mode [Silent/Normal] & On change & Small \\
\hline
21 & Running Applications & Once every 5 seconds & Small \\
\hline
22 & Screen Status [ON/OFF] & On change & Small \\
\hline
23 & Step Counter & up to 10 samples per second & Small \\
\hline
24 & Step Detection & On change & Small \\
\hline
25 & Touch event & On change & Small \\
\hline
26 & User Presence & On change & Small \\
\hline
27 & WIFI Network Connected to & On change & Small \\
\hline
28 & WIFI Networks Available & Once every minute & Small \\
\hline
\end{tabular}
\label{sw_sensors}
\end{table}

%\vspace{-0.2cm}
\begin{table}[htb!]
\centering
\caption{QA sensors}
\begin{tabular}{|c|l|l|l|}
\hline
\textbf{No} & \textbf{QA Sensor}             & \textbf{Estimated Frequency} & \textbf{Category} \\
\hline
29 & Time Diary question        & On change           & Small    \\
\hline
30 & Time Diary confirmation     & On change           & Small    \\
\hline
31 & Time Diary answer          & On change           & Small    \\
\hline
32 & Task question              & On change           & Small    \\
\hline
33 & Task confirmation           & On change           & Small    \\
\hline
34 & Task answer                & On change           & Small    \\
\hline
\end{tabular}
\label{questionnaire_sensors}
\end{table}
\noindent
Differently from HW and SW sensors, QA sensors are specific to iLog and, as far as we know, are not found in any other data collection APP. They are the key element which enables the design and implementation of the scheduling language iLogCal described below and, therefore, of the entire data collection methodology described in this paper. Looking at Table \ref{questionnaire_sensors}, it is possible to notice two sets of QA sensors. The first set, concerning \textit{Time Diaries}, is used to answer questions about the context of the experiment, while the second set, concerning additional \textit{Tasks}, is used to get information about the data collection process. As an example of task, a user may be asked whether it confirms a previous answer, or if it achieved a specific task, e.g., returning back a missed phone call).
Inside each set of QA sensors we have three types of sensors, as follows:

\begin{itemize}
    \item 
Time Diary / Task question: when a question is generated, ready for delivery;
\item 
Time Diary / Task confirmation: when a
question, is delivered to the device of the
participant (who may then look at it in any moment after this);
\item  Time Diary / Task answer: when an answer is stored, with  additional information of the difference between answer and notification
time (the notification time and time defined by the researcher when the question is to be submitted to the participant).
\end{itemize}
\noindent 
As described in detail in \cite{bison2024impacts}, the \textit{reaction time}, also called the \textit{response time}, that is, the time difference between when one receives a question and when (s)he starts filling the answer, and the \textit{notification time},  also called the \textit{completion time}, that is, the time taken to fill an answer, are key factors which impact the quality of an answer. Here, by the quality of an answer, we mean an answer which has been meaningfully provided (and not just dropped) and which is correct. It can be noticed that response time and completion time can be easily computed, for each question and (possibly missing) corresponding answer, from the information provided by QA sensors.

The work in \cite{bison2024impacts} also provides evidence of the fact that response and completion time are highly impacted by the situational and temporal contexts. iLog has various features which allow; (i) to compute this information and, even more importantly, (ii) to provide extreme flexibility about the configuration of the data collection. These two features are key, together with QA sensors, for learning about which factors influence the user behavior. Let us consider some examples. 
\textit{First}, the researcher has a wide variety of question types to select from. This can be exploited to ask questions which are not related to the data to be collected as part of the experiment as such, but which are about the data collection process meta-context. For instance, as done in \cite{KD-2019-Zeni-UBICOMP}, under certain conditions, the participant can be asked to confirm a specific answer. 
\textit{Second}, the possibility of configuring the data collection frequency independently for each sensor allows to collect data whose main purpose is solely that of monitoring the experiment evolution. As an example, collecting the GPS when asking a question about the current location allows iLog to validate the correctness of the answer. This idea is exploited in the work described in \cite{KD-2018-Giunchiglia-PERCOM1,KD-2019-Zeni-UBICOMP}. The same applies to the bluetooth or to any other sensor which provides information about the situational context of the question-answering process. As another example, any question and relative answer about the current situational context is key for collecting information about the meta-context at the precise moment when an answer is provided. This idea is exploited in the work described in 
\cite{bison2024impacts} for computing the best moment for asking a question.
\textit{Third}, the information provided by SW sensors, if integrated with the information about the experiment temporal context, allows to understand and correlate the activities performed by participants. For instance, \cite{KD-2017-SOCINFO} exploits this information to correlate academic performance and social media usage, while \cite{kasinidou2024artificial} uses this information to detect the usage of social media during lectures.

As a conclusive remark, it is worth noticing that the amount of personal information that has been collected in this experiment and that, in general, which can be collected using iLog, is huge, thus raising important privacy and ethics related issues. The approach that we follow is based on tree main pillars. The first is that the use of iLog follows a very precise GDPR and ethics aware methodology, inspired by the approach followed in
\footnote{https://aihumanimpact.org/}. The details of how this was applied in the experiment described in this section are reported in \cite{bussoDiversityOne}. The second is that our main focus is on research-motivated data collections. The third is that iLog is being redesigned to store and keep all the data in the participant's device. Ultimately, in the next version of the platform, the participant will be in full control of the data and of how to use them for his/her own purpose. In turn this will allow for the possibility, at the moment  unexplored, of increasing the participant's \textit{self-awareness} of his/her life-style and habits, as modeled as (life) sequences of situational contexts (see Section \ref{repctx}) and visualized by the dashboard (see Section \ref{monitoring}). Some early ideas in this direction are provided in \cite{li2022representing}.

%% file: section/3-Representing_context.tex
\begin{figure}[!ht]
    \centering
\includegraphics[width=0.75\textwidth]{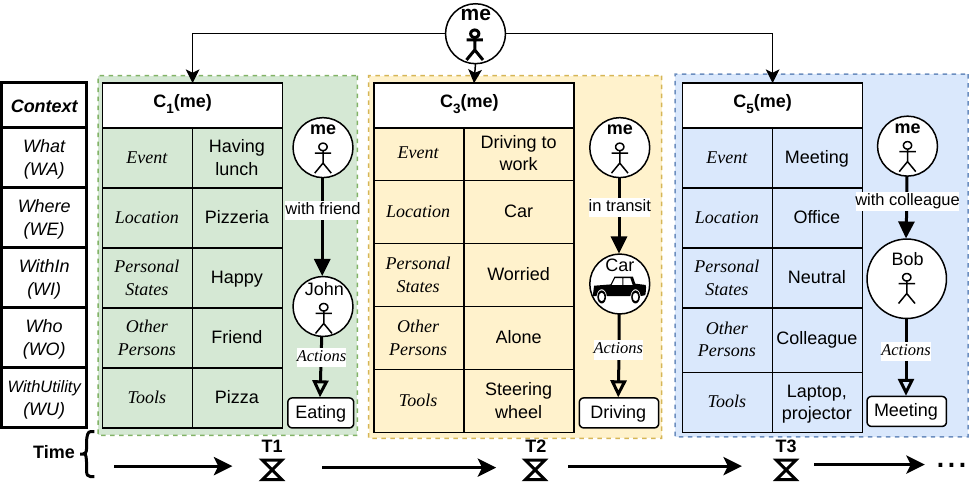}
\caption{An example of everyday life sequence.}
\label{fig:context-everyday-life}
\end{figure}

%\begin{figure}[htp!]
%\centering
%\includegraphics[scale=0.6]{images/context.drawio.png}
%\caption{\centering The situational context model.}
%\label{Context}
%\end{figure}

\section{Representing the Situational Context}
\label{repctx}

 The notion of context used here is an elaboration of the notion of context first introduced in \cite{giunchiglia2017personal} and further extended in \cite{KD-2022-Xiaoyue}. As a motivating example, let us consider a small portion, of the duration of around a couple of hours, of the everyday life of the students participating in the experiment described in Section \ref{study}, as represented in Fig. \ref{fig:context-everyday-life}. Let us call this person, \textit{me}. As from Fig. \ref{fig:context-everyday-life}, the activities of \textit{me} consist of the following:
 
\begin{itemize}
    \item During a first period of time T1 (green box), \textit{me} is at a pizzeria having lunch with the friend, John. They are having pizza and \textit{me} is happy;
    \item Then, in the following period of time T2 (orange box), \textit{me} is driving to work, alone and  is in a worried mood;
    \item Finally, during T3 (blue box), \textit{me} is in a meeting in office with the colleague Bob and \textit{me} is in a neutral mood.
\end{itemize}
Following the terminology introduced in \cite{KD-2022-Xiaoyue},
what is represented in  Fig. \ref{fig:context-everyday-life} is a specific instance of a (small) fragment of the life of  \textit{me}, written as the \textit{Life Sequence} of \textit{me, $L(me)$}, that is, a sequence of three \textit{situational contexts} of \textit{me}, written $C_i(me)$ for a total time duration of T1+T2+T3. We define Life Sequences as follows
  
\vspace{-0.7cm}
\begin{center}
\begin{equation}
 \hspace{3cm}  
L(me) = \langle C_1(me), \dots, C_n(me) \rangle \ \ \ \ \  \text{with} \ \ \ 1 \leq i \leq n
\label{eq:ls}
\end{equation}
\end{center}

\noindent 
We assume that \textit{me} is involved in only one personal context at the time. This models the intuition that a situational context is associated to a single location, that is, that moving from one location to another means changing context, and that, at any given moment, a person can be in only one place. A life sequence fully covers the period under consideration, but there may be elapsed times between a context and the next one in the sequence. What makes a set of contexts a life sequence is not the time sequentiality but the fact that they are functionally related by some overall motivation or purpose. Some examples of life sequences are: the lectures in a morning, which may or may not have an elapsed time in between, depending on whether the class is in different rooms, where two classes in the same room can be modeled as a single context or as the sequence of two contexts in the same location; the lectures of the same course in a semester; the editions of the same course along five years; a portion of everyday life as in Fig. \ref{fig:context-everyday-life}; a full day, and so on.

We model a situational context in terms of five components as follows (from now on we drop the argument \textit{me} whenever no confusion arises).

\vspace{-0.8cm}
\begin{center}
\begin{equation}
 \hspace{4cm}   
 C = \langle WE , WA ,  WI , WO , WU \rangle
\end{equation}
\end{center}

\begin{figure}[htp!]
\centering
\includegraphics[scale=0.65]{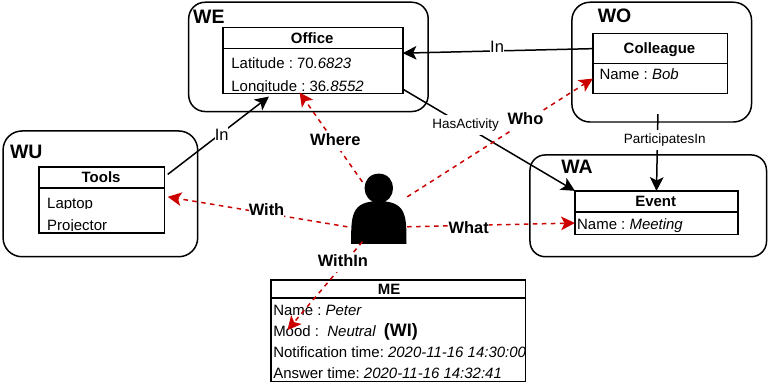}
\caption{\centering The knowledge Graph of the third situation context in  Fig. \ref{fig:context-everyday-life}.}
\label{example}
\end{figure}

\vspace{-0.3cm}
\noindent
where:
\begin{itemize}
\item \textit{WE}, the so-called \textit{spatial context}, is a linguistic description, e.g., a label or some text provided in a formal or natural language, describing the \textit{location} where \textit{me} is at the moment. Information about it can be obtained from the sensor data as well as from iLog questions. In the experiment described in Section \ref{study}, the HW sensors that can be used to compute the location are, e.g., GPS or WI-FI. The name of 
the spatial context is the label (selected from a set of predefined ones) provided by the answer to the question  “\textit{\textbf{W}h\textbf{E} are you?}”. In the third context of Fig. \ref{fig:context-everyday-life}, the place where \textit{me} is located is an \textit{office}.

\item \textit{WA}, the so-called \textit{activity} or \textit{event context}, is a linguistic description of the \textit{activities} being currently performed by \textit{me}. A single context may contain one or more activities which in turn, can be performed in sequence or in parallel \cite{li2022representing}. Information about this can be obtained from  sensor data as well as from the iLog questions. In the experiment described in Section \ref{study}, a HW sensor that can be used to know about the physical activities is, the accelerometer; whereas, a SW sensor  can be used to know about the online activities a person is doing, while the QA sensors allows to know when \textit{me} is involved in which question-answering activities. The name of 
the temporal context is the label (selected from a set of predefined ones) provided by the answer to the question  “\textit{\textbf{W}h\textbf{A}t are you doing?}”. In the third context of Fig. \ref{fig:context-everyday-life}, the temporal activity being carried out is a \textit{meeting}.

\item \textit{WI}, the so-called \textit{internal (activity} or \textit{event) context}, is a linguistic description describing the \textit{internal activities} occurring inside \textit{me}. Information about this can be obtained from the sensor data (e.g., heart beat, blood pressure) provided by medical devices or smart watches, as well as from the iLog questions. In the experiment described in Section \ref{study}, no sensor could provide this type of information. The only question providing this type of information was the question asking "\textbf{W}hat mood are you \textbf{I}n?".  
In the third context of Fig. \ref{fig:context-everyday-life}, the mood of \textit{me} is \textit{neutral (agreeable)}.

\item \textit{WO}, the so-called \textit{social context}, is a linguistic description describing the \textit{people}, possibly none, who are with \textit{me} at the moment. Information about it can be obtained from sensor data as well as from iLog questions. In the experiment described in Section \ref{study}, the HW sensors that can be used to compute the social context are, e.g., the blue-tooth or the microphone (via speaker recognition). The social context is described by the
 label (selected from a set of predefined ones) provided by the answer to the question  “\textit{\textbf{W}h\textbf{O} are you with?}”. In the third context of Fig. \ref{fig:context-everyday-life}, \textit{me} is with one or more \textit{colleagues}.

 \item \textit{WU}, the so-called \textit{material} or \textit{(tool / utensil) context}, is a linguistic description, describing the \textit{tools}, possibly none, which are used or usable by \textit{me}. Example of tools are: the car used in a trip or the phone used in the interaction with a friend. In the experiment described in Section \ref{study}, no sensor and no question was used to provide this type of information.
It could be obtained from sensor data (e.g., bluetooth, rfid, wifi) as well as from a question like “\textit{\textbf{W}hich \textbf{U}tensils are you using?}”. In the third context of Fig. \ref{fig:context-everyday-life}, \textit{me} is using a few objects, e.g., a \textit{projector}, and a \textit{laptop}.
\end{itemize}
\noindent
The scenario in  Fig. \ref{fig:context-everyday-life} can be modeled by a Knowledge Graph (KG) \cite{giunchiglia2023context}, see 
Fig. \ref{example}. We can identify the following components :
\begin{itemize}
\item a (sub-)KG for each context, including the internal context;

    \item a \textit{node} for each entity involved, e.g., \textit{Person}, \textit{Room}, and \textit{Furniture};
    
    \item an \textit{attribute} and corresponding \textit{value} for each node / entity; for instance, the \textit{attributes} of \textit{ME} (whose context we are describing) are \textit{Name}, \textit{Mood}, \textit{Notification time} and \textit{Answer time};
    
    \item an \textit{Edge} for each relation between two entities; for instance, \textit{office} is $PartOf$ $work place$, whereas; \textit{Bob (person)} and the \textit{Office table (Furniture)} are both $in$ the \textit{office} where a $meeting$ taking place.
\end{itemize}
% Each node represents an entity, e.g., \textit{person} and \textit{room}, with
% their respective attributes with values; for instance, attributes of \textit{ME (whose context we are describing)} are \textit{Name}, \textit{Mood}, \textit{Notification time} and \textit{Answer time} with the corresponding values as shown in Fig. \ref{example}. Edges represent relations between entities, e.g., \textit{office} is $PartOf$ $work place$, whereas; \textit{Bob (person)} and the \textit{Office table (Furniture)} which are both $in$ the \textit{office} which $HasActivity$ of a $meeting$ taking place.
\noindent
A Life Sequence is represented as a set of KGs. 
As it will be described in detail in the next section, any of the attributes and relations occurring in one or a combination of KGs can be used as a precondition enabling or disabling a question or a task of iLog.

%% file: section/4-Scheduling_plan.tex
\section{Representing the temporal context}
\label{scheduling1}

We model the temporal context using iLogCal, a scheduling language developed on top of iCal,
the iCalendar/RFC5545 (iCal) standard. Using the iCal standard provides three primary benefits, namely; 
\begin{itemize}
    \item An explicit and declarative representation of the activities involved in an experiment. This, in turn, allows to easily modify, for instance via a graphical interface, the resulting schedule, e.g., for instance by eliminating questions or adjusting response times; 
    \item The possibility of using the iCal Recurrence Rule (RRule) for the formulation of recurring activities;
    
    \item Access to advanced open-source graphical interfaces in the form of calendar-like visualizations, e.g., \textit{Fantastical Calendar} \footnote{https://flexibits.com/fantastical}.    
\end{itemize}
\noindent
iLogCal organizes the specification of an experiment in three main components, as follows:
\begin{enumerate}
    \item The general schedule aggregating all the different components;
    \item The question answering component;
    \item The sensor data component.
\end{enumerate}
We introduce and discuss below a snapshot of the three iLogCal components using the Extended Bachus-Naur Form (BNF) notation. We use the following conventions: terminal symbols are written in the font of the text of the paper, \texttt{<text>} is a nonterminal symbol for any \texttt{<text>}, \texttt{<text1><text2>} is the sequence of \texttt{<text1>} and \texttt{<text2>}, \texttt{<text1>|<text2>} stands for a choice between \texttt{<text1>}  and\texttt{<text2>}.
\texttt{\{<text>\}} is any number of occurrences of \texttt{<text>}, allowing for zero occurrences.

\begin{figure}[htbp!]
\vspace{-0.5cm}
\centering
\begin{xltabular}{\linewidth}{l r X}
\bnfterm{User} &\bnfdef& \{\bnfterm{Calendar}\}\bnfend 
\bnfterm{Calendar} &\bnfdef& \bnfterm{Calendar id} \{\bnfterm{Context collection}\} \bnfend 
\bnfterm{Calendar id} &\bnfdef& Integer \bnfend
\bnfterm{Context collection} &\bnfdef& \{\bnfterm{Question collection}\} 
\{ {\bnfterm{Sensor collection} \}} \bnfend \\
\end{xltabular}
\caption{ \centering Experiment General schedule.}
\label{BNF1}
\end{figure}

\subsection{Experiment General schedule}

The BNF of an experiment general schedule is reported in Fig,\ref{BNF1}. We have the following
 observations. 

\begin{itemize}
    \item A user may be associated with multiple calendars; this allows a user to participate in multiple experiments in parallel. A life sequence consists of the data collected by one of more calendars; 
    \item Each calendar contains multiple context collections; this allows to have an articulated specification of an experiment, while maintaining the unity of the same experiment;
    \item Each context collection allows for any number of 
    question collections as well as  sensor collections. This facilitates the specification of multiple diverse data collections within the same context collection. 
    \item Identifiers, e.g., \texttt{<calendar id>} allow to identify components inside the same group from one another. The full version of iLogCal has ids for each and every element listed in Fig. \ref{BNF1}.
\end{itemize}
As an example of instantiation of the schedule in Fig.\ref{BNF1}, the experiment described in Section \ref{study} was organized as follows. Three calendars were created to collect data from participants. The first calendar gathered general questions about the day, the second calendar collected time diary questions as detailed in Section \ref{study}, and the third calendar compiled additional questions regarding food and drinks. Notably, all three calendars gathered the same sensor data. The researcher then decided how to use the sensor data collected from the different calendars. The details of this step are discussed in the following subsections.

\subsection{Question collection}

The BNF of the iLogCal sensor collection is reported in Fig,\ref{BNF2}. We have the following observations. 
\begin{itemize}
    \item A \texttt{<Question collection>} is associated with;  a unique id \texttt{<Cid>}; a time range specified by a start and end time (\texttt{<start>}, \texttt{<end>});   a variable \texttt{<status>} which indicates whether the event has been accepted by the user, with 1 meaning acceptance and 0 meaning rejection; a Recurrence Rule \texttt{(<RRule>)} and a \texttt{<question>};
    
    \item \texttt{<RRule>} specifies the values used to determine each recurrence and how the event should be repeated. It includes three elements: \texttt{<Interval>}, \texttt{<Count>}, and \texttt{<Frequency>}. \texttt{<Interval>} specifies the number of \texttt{<Frequency>} units that must elapse before the next occurrence of the event. \texttt{<count>}  specifies the number of times the event will be repeated;
    
    \item Each \texttt{<question>} encompasses a question category \texttt{<Qcategory>}, as defined in section \ref{repctx}. \texttt{<question content>} consists of the text of the query to be posed to the participant, e.g., “\textit{What are you doing?}”. The  participant answers,  e.g., “\textit{Meeting}”, are included in  \texttt{<answer content>}. The field \texttt{<Qtype>} is instantiated following, with minor adaptations, the state of the art from the Social Sciences, see, e.g., \cite{bradburn2004asking}. The values written in the BNF are those used in the experiment in Section \ref{study}.
    \item We have five possible values for \texttt{Qcategory}, one for each type of context, as from above. That is, we associate a question collection to each any relevant context type. Given that, as from above, we may have any number of question collections we may have zero, one or multiple instances of the same context type.
    \end{itemize}

\begin{figure}[htbp!]
\vspace{-1.0cm}
\centering
\begin{xltabular}
{\linewidth}{l r X}
\bnfterm{Question collection} &\bnfdef&  \bnfterm{Cid} \bnfterm{start} \bnfterm{end} \bnfterm{status} \bnfterm{RRule} \bnfterm{question}\bnfend
\bnfterm{Cid} &\bnfdef& Integer \bnfend
\bnfterm{dtstart} &\bnfdef& DateTime \bnfend
\bnfterm{dtend} &\bnfdef& DateTime \bnfend
\bnfterm{status} &\bnfdef& Boolean \bnfend
\bnfterm{RRule} &\bnfdef&  
\bnfterm{Interval} \bnfterm{Count} \bnfterm{Frequency} \bnfend
\bnfterm{Interval} &\bnfdef& Integer \bnfend
\bnfterm{Count} &\bnfdef& Integer \bnfend
\bnfterm{Frequency} &\bnfdef&  
Daily \bnfor Weekly \bnfor Monthly \bnfor Yearly \bnfend
\bnfterm{question} &\bnfdef&  \bnfterm{Qid} \bnfterm{Qcategory} \bnfterm{Qcontent} \bnfterm{Qtype} \bnfend
\bnfterm{Qid} &\bnfdef& Integer \bnfend
\bnfterm{Qcategory} &\bnfdef& WE \bnfor  WA \bnfor  WI \bnfor WO \bnfor WU\bnfend
\bnfterm{Qcontent} &\bnfdef&  \bnfterm{question content} \bnfterm{answer content} \bnfend
\bnfterm{question content} &\bnfdef& String \bnfend
\bnfterm{answer content} &\bnfdef& String \bnfend
\bnfterm{Qtype} &\bnfdef& Dichotomous Question \bnfor Multiple choice question \bnfor Single choice question \bnfor Free Text Question \bnfend
\end{xltabular}
\caption{ \centering Question collection.}
\label{BNF2}
\end{figure}

\noindent
    We illustrate the use of the BNF in Fig.\ref{BNF2} by applying it to the  data in Fig. \ref{example}. We select WA as the value for \texttt{Qcategory}. As a consequence we have that the value for \texttt{question context} is “What are you doing?”. We set the \texttt{Qtype} as a single choice question and we have \texttt{answer content} instantiated to “meeting”, i.e., the input from \texttt{me}. As from Section \ref{study}, during the first two weeks questions were dispatched 48 times a day (once every 30 minutes) over a span of 14 days (two weeks). This was achieved by setting an \texttt{RRule} with a \texttt{<Frequency>} set to daily, an \texttt{interval} of 48 times, and a \texttt{count} of 14.

\subsection{Sensor data collection}

The BNF of the iLogCal sensor data collection is reported in Fig,\ref{BNF3}.  The structure is essentially the same as that used for question collections and exploits a similar set of nonterminal symbols. \texttt{<Name>} is the name of the specific sensor, e.g., GPS, accelerometer, \texttt{<Description>} explains how the specific sensor should be used;  \texttt{<Sensor type>}, identifies the family to which the specific sensor belongs. For instance the GPS is a Location sensor, while the accelerometer is a Motion sensor. 
We illustrate the BNF in Fig.\ref{BNF3} by applying it to to the collection of the GPS data in Figure \ref{example}, as part of the collection of data of the user’s spatial context (WE). In this case, the field \texttt{Description} is filled with, e.g., a text saying \textit{"Location information using GPS connections"}. Furthermore, as from Table \ref{hw_sensors}, the GPS information was collected every minute for 48 days; this means a \texttt{RRule} with \texttt{<Frequency>} set to Minute,  \texttt{interval} to 1, and a \texttt{count} to 69120 (= 48 x 24 x 60).

\begin{figure}[htbp!]
\vspace{-0.8cm}
\centering
\begin{xltabular}
{\linewidth}{l r X}
\bnfterm{Sensor Collection} &\bnfdef&  \bnfterm{Sid} \bnfterm{start} \bnfterm{end} \bnfterm{RRule}, \bnfterm{sensor} \bnfend
\bnfterm{Sid} &\bnfdef& Integer \bnfend
%\bnfterm{dtstart} &\bnfdef& DateTime \bnfend
%\bnfterm{dtend} &\bnfdef& DateTime \bnfend
%\bnfterm{sensorRRule} &\bnfdef&  
%\bnfterm{Interval}, \bnfterm{Count}, \bnfterm{Sfrequency} \bnfend
%\bnfterm{Interval} &\bnfdef& Integer \bnfend
%\bnfterm{Count} &\bnfdef& Integer \bnfend
\bnfterm{Frequency} &\bnfdef&  
Millisecond \bnfor Second \bnfor Minute \bnfor Hour \bnfend
\bnfterm{sensor} &\bnfdef&  
\bnfterm{Name} \bnfterm{Description} \bnfterm{Sensor type} \bnfend
\bnfterm{Name} &\bnfdef& String \bnfend
\bnfterm{Description} &\bnfdef& String \bnfend
\bnfterm{Sensor type} &\bnfdef& Social \bnfor Motion \bnfor Location \bnfor Inertial \bnfor Device \bnfor Ambient\bnfend
\end{xltabular}
\caption{ \centering Sensor data collection.}
\label{BNF3}
\end{figure}

%% file: section/5-Monitoring_plan_execution.tex
\begin{figure}[h!]
\centering
\includegraphics[scale=0.55]{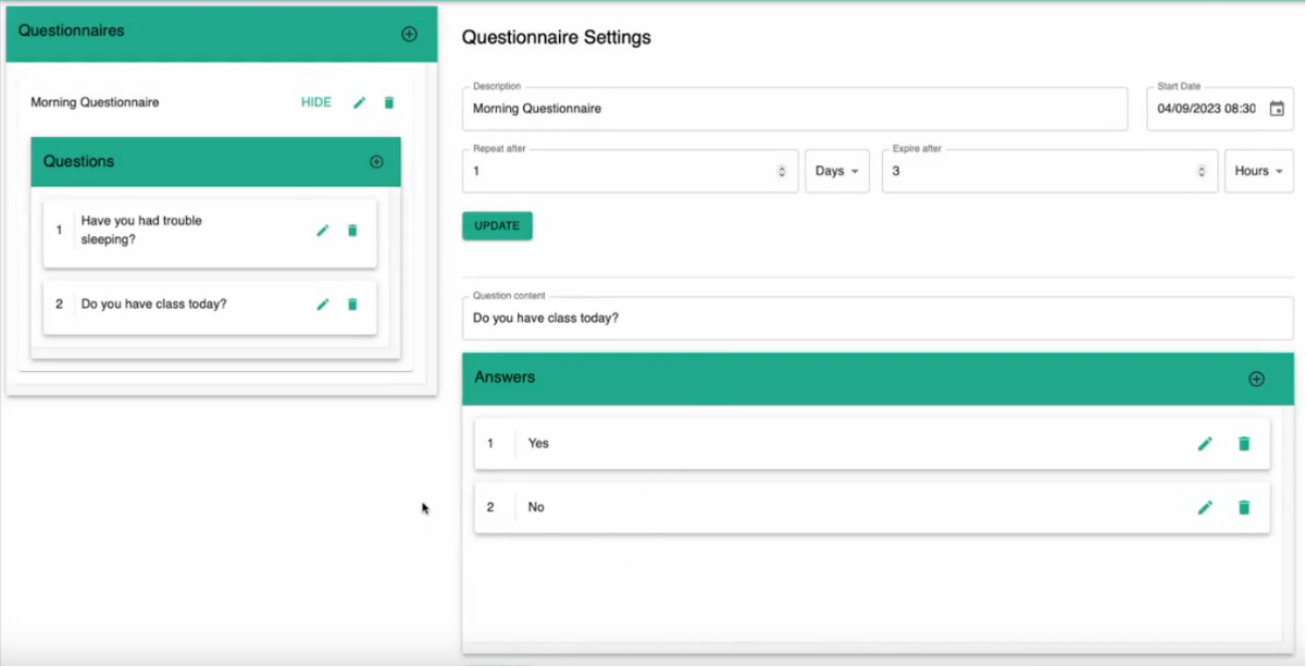}
\caption{\centering Using ISAC to generate a question.}
\label{fig:isac_example}
\end{figure}

\newpage
\section{Monitoring the Data Collection}
\label{monitoring}
\begin{figure}[h!]
\centering
\includegraphics[scale=0.4]{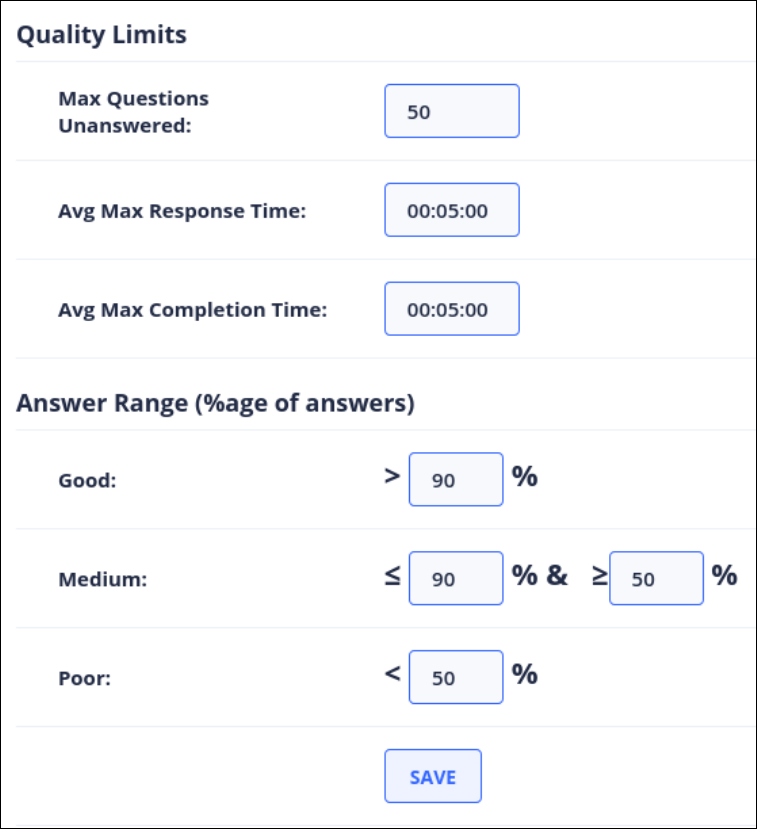}
\caption{\centering Quality parameters used to rank participants.}
\label{fig:quality_lim}
\end{figure}

Monitoring the data collection entails tracking how well an experiment is executed based on a comparison with a predefined experiment plan. The first main component for experiment control is \textit{ISAC (iLog System Administration Component)}, a tool which allows for the creation of the experiment plan/calendar. ISAC allows researchers; 
(1) to create the experiment plan;
    (2) to visualize the experiment Timeline where Researchers can see the entire study schedule at a glance, making it easier to plan and adjust various phases of the data collection process; and
    % \item Set Reminders and Notifications: Automated reminders and notifications can be scheduled to prompt participants to complete questionnaires or ensure their devices are collecting sensor data as required.
    (3) to adapt the sampling frequency dynamically, thus adjusting the intervals at which data is collected or questionnaires are sent out.
Figure \ref{fig:isac_example} depicts ISAC when used to generate a question in the definition of the experiment plan. The information necessary includes all the elements of the BNF defined in the previous section and, in particular,   name and description of the question, repetition frequency,  available answer options, and scheduled day and time for sending it out.
 As a complement to ISAC, the platform features a component where the researcher can also set various parameters which then are use to rank the quality of the participants' involvement in the experiment, see Figure \ref{fig:quality_lim}. These parameters include: maximum allowed number of unanswered questions, average maximum completion and response time. During the execution of the experiment, a participant's performance can be ranked as $good$, $medium$ or $poor$. In case of $poor$ behavior, the participant may be kicked out of the experiment and, for instance, not get the economic incentive if this was promised (as it was the case in the experiment in Section \ref{study}).

% \begin{figure}[h!]
%     \centering
% % \includegraphics[scale=0.5]{images/Figure 6.pdf}
% \caption{Filtering of participant answers per day. At the top are filtering controls while at the bottom is a heatmap with the x-axis representing dates and the y-axis participants. Green represent high values and orange, low values}
% \label{fig:dashboard_filter}
% \end{figure}
\noindent
The second main component is a \textit{dashboard} which % , see Fig. \ref{fig:sample_summary} for its main interface, the one which is presented to the user when first logging in. The dashboard 
enables researchers and participants to efficiently track the progress of the experiment execution. It provides real-time insights and control, ensuring that researchers  maintain the quality and integrity of the data collection process while also supporting the participant engagement and compliance. 
% Both are able to browse and evaluate the data collected and perform different queries on this data instantly.The module allows researchers to monitor data collection in real time. 
This includes tracking both sensor data and responses to questions made by participants. Key features include:
\begin{itemize}
    \item \textit{Live Data Feeds}. This module allows researchers to view the incoming data as it is collected, providing immediate visibility of the participant activity and data trends in real-time.
    
    \item \textit{Compliance Tracking}. This module displays participant compliance rates, showing who is completing the required tasks (answering questions) and who might be falling behind. This allows researchers to quickly identify and address potential issues.
    \item \textit{Data Quality Checks}. This module consists of a set of algorithms which discover  possible participant misbehavior and/ or errors in the data. This is used to notify researchers  and also participants so that they can take corrective actions promptly.
    \item \textit{Advanced Analytics}. It incorporates data filtering techniques that enable researchers to conduct real-time data analysis, swiftly discovering trends and creating insights. 
    % For instance, Figure \ref{fig:dashboard_filter} depicts a researcher's application of filters to the collected answers from participants.
    % \item MISSSING 1
    % \item MISSING 2
\end{itemize}
\begin{figure}
        \centering        \includegraphics[width=1\linewidth]{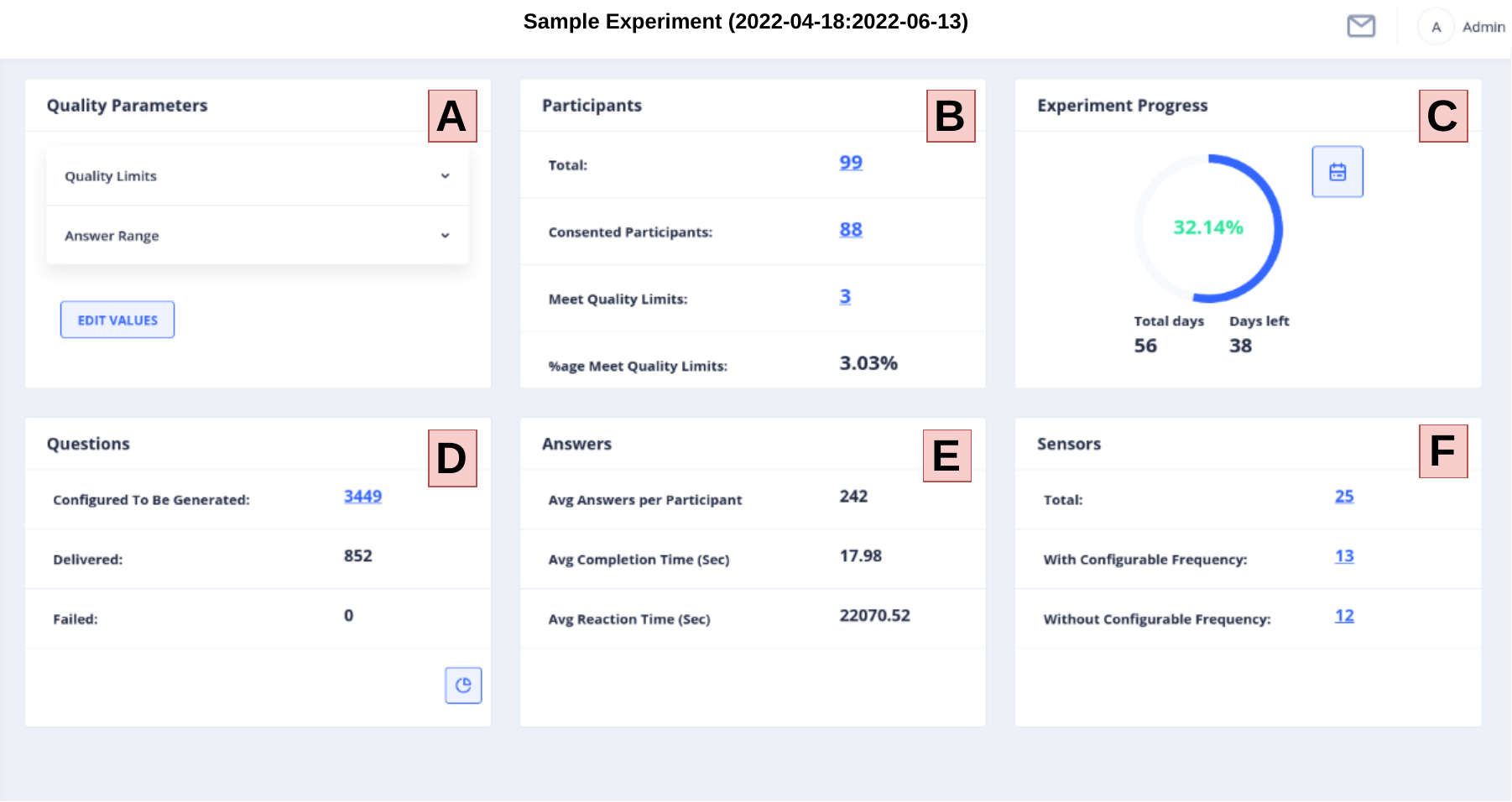}
        \caption{Dashboard: summary of an experiment}
        \label{fig:sample_summary}
    \end{figure}
The main interface of the dashboard is reported in Fig. \ref{fig:sample_summary}. This is the main interface presented to the user when (s)he logs in. In  \ref{fig:sample_summary} we can identify (left to right, top to bottom) the following elements:
\begin{enumerate}[label=\Alph*.]
    \item Here, the user is presented with the \textit{quality parameters} set by the researcher (see Fig. \ref{fig:quality_lim}). The researcher can modify them by calling the module mentioned above, while the participant can only view them.
    \item This section is a summary of the number of participants in the experiment in real time. This section is visible only to the researcher.
    \item This section reports the progress of the experiment, in terms of the number of days covered or left.
    \item Question delivery is key when monitoring an experiment. This section shows the level at which questions are being delivered to the user.
    \item For any experiment, the number of answers given affects its overall quality. The researcher is presented with an average of all the answers in the experiment, whereas participants view a summary of their answers.
    \item As with questions and answers, the sensors section helps the user understand the sensors being collected. The frequency of collection is also reported.
\end{enumerate}

\begin{figure}[htp!]
\centering
\includegraphics[scale=0.38]{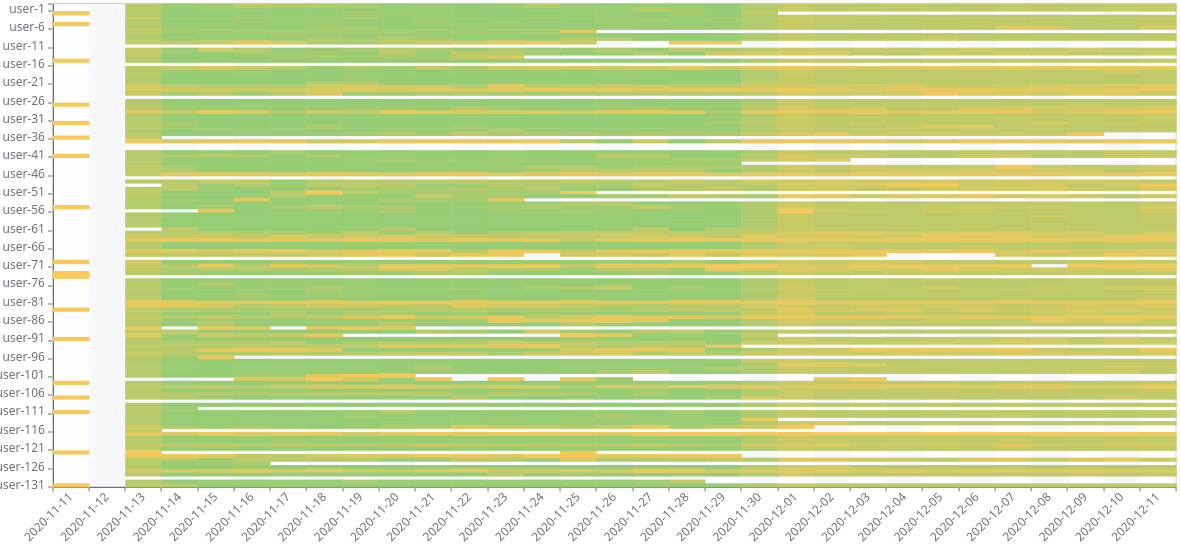}
\caption{ \centering Heatmap of participants' answers per day}
\label{particiant_answers}
\end{figure}
\begin{figure}
    \centering    \includegraphics[width=0.7\linewidth]{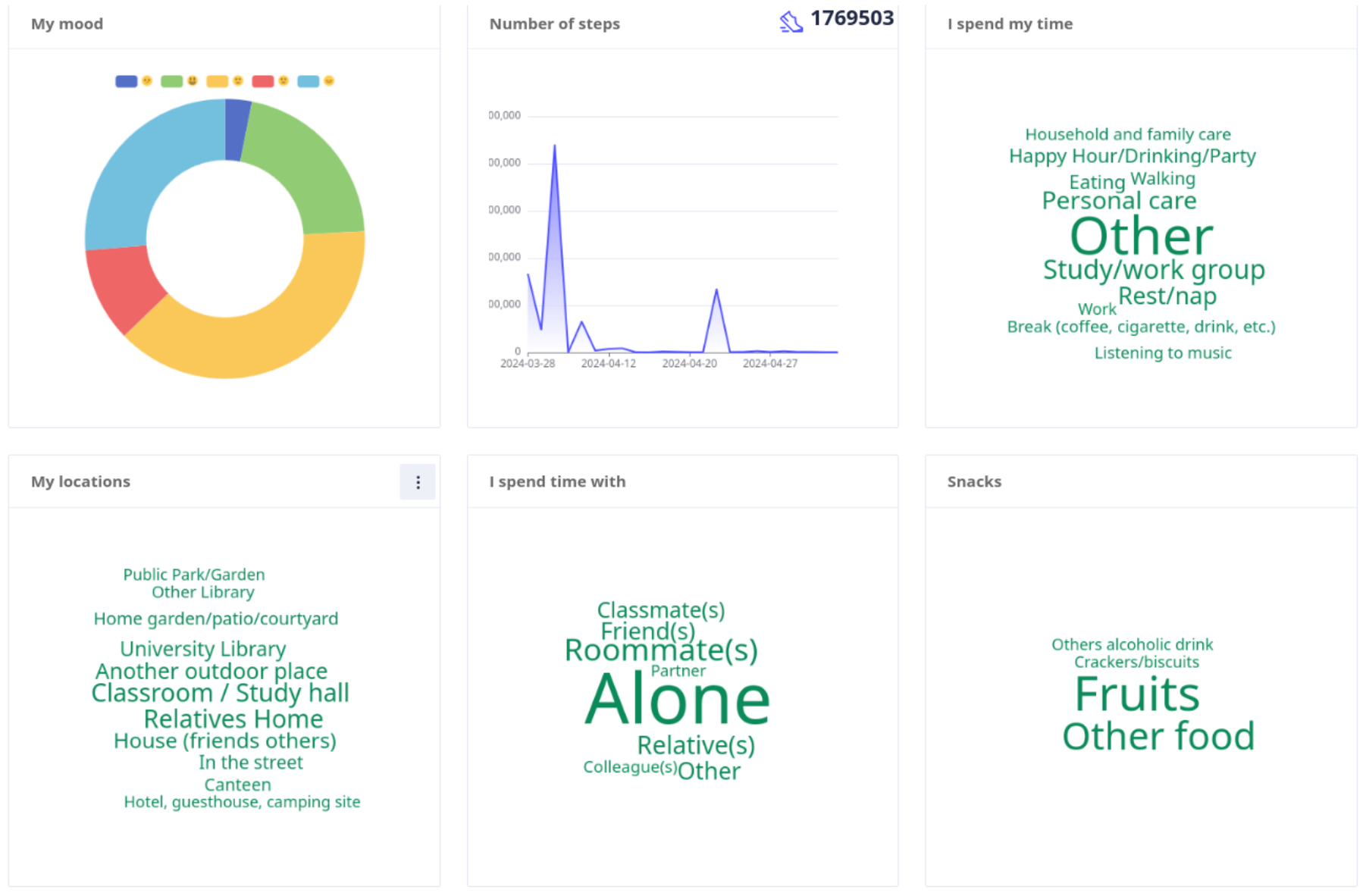}
    \caption{A participant's data as seen from the Dashboard.}
    \label{fig:participant_data}
\end{figure}
\begin{figure}[htp!]
\centering
\includegraphics[width=0.9\linewidth]{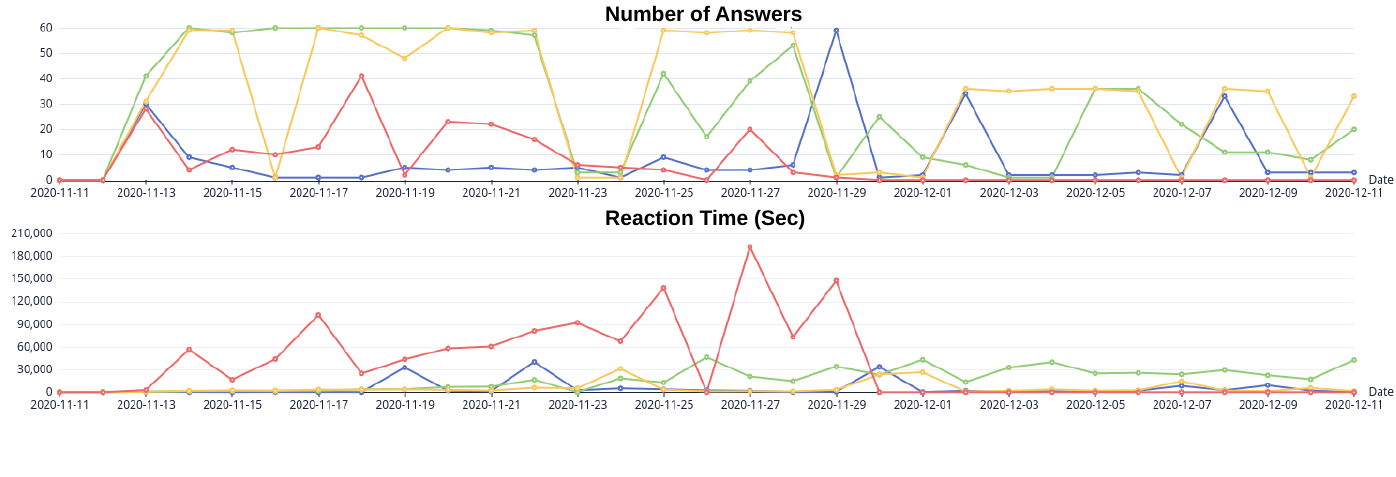}
\caption{ \centering Comparison of answering behavior of three other participants}
\label{compare_participant_answers}
\end{figure}

\noindent
The dashboard main view is integrated with various other additional visualizations which focus on specific aspects of the data collection. For instance, Fig. \ref{particiant_answers} displays a heatmap of the participants' responses from each experiment day, with green denoting participants with a high answer rate and yellow denoting those with a low answer rate. Participants are represented on the x-axis, while experiment days selected on the y-axis. Looking at the heatmap closely, it can be noticed that no data was recorded for the date \textit{'2020-11-12'}, which may indicate a systemic issue that needed to be fixed before the experiment further progressed successfully. As another example, Fig. \ref{fig:participant_data} reports the outcome of a functionality which allows the researcher, but also the single participants, in this latter case restricted to their own personal data, to explore and navigate their collected data.
Based on the questions and nature of the experiment, they can learn about their lifestyle, such as whether they eat a lot of snacks, spend a lot of time in one place, or even experience mood swings. As a last example, in Fig. \ref{compare_participant_answers}, three participants (orange, red, and blue lines) are compared with another participant (green line). Two graphs are displayed; the top showing the answers contributed so far; and the one at the bottom, the time in seconds before starting to answer (so called, reaction time). 

%% file: section/6-Learning.tex
\section{Improving the answer quality}
\label{learning}

We are interested in collecting context information via smartphone questions. However, these questions,  when asked frequently, can become intrusive, especially when they interrupt users during periods of activity or when in mobility. As a consequence,
the quality of the answers from humans is not always as high as needed. 
 Users often do not read, or do not answer, or provide wrong answers to machine-asked questions, or turn-off their data collection APP, and more; see, e.g., \cite{bison2024impacts,KD-2020-Bontempelli1}. This problem can arise from various factors \cite{furnham1986response}, such as recall bias, where participants do not accurately recall previous activities \cite{porta2014dictionary}, missing or incorrect responses \cite{H-2015-Schneier,bison2024impacts}.
And it becomes particularly acute when one tries to scale the collection of big-thick data to (life) long, human-in-the-loop human-machine interactions \cite{KD-2022-Bontempelli-lifelong}, that is, the applications which motivate the work described in this paper. To address these challenges, it is essential to develop methodologies that not only optimize the timing of questions to minimize interruptions but also improve the overall quality of the responses collected. 

By answer quality, we focus here on the number of correct answers, rather than on the number of missing answers. As already mentioned in Section \ref{study}, the state of the art suggests that the quality of answers is influenced by reaction time \cite{H-2019-Van, KD-2023-Bison-answer, bison2024impacts}. The shorter the reaction time, the higher the quality of answers, a factor also aligned with the recall bias theory. The aim of this section is to describe a ML component, exploiting the notions of situational and temporal context described above, capable of learning when to ask a question so that to minimize the reaction time. The proposed ML algorithm exploits the following information:

\begin{itemize}
\item \textit{Temporal context.} We consider the day of the week, represented numerically from 1 (Monday) to 7 (Sunday) plus the 
 \textit{Hour of the day.} organized in four time periods: Morning (6 AM to 11 AM), Afternoon (12 PM to 5 PM), Evening (6 PM to 11 PM), and Night (12 AM to 5 AM).

\item \textit{Situational context.} We consider  the answers to the three questions \textit{"Where are you?"} (the spatial context), \textit{"What are you doing?"} (the activity context),  and \textit{"Who are you with?"} (the social context). 

\item \textit{Demographics.} We consider information is consistently used to characterize individuals through ascriptive and acquisitive traits \cite{blau1967american}. Namely, we used the gender, degree and department of each participant.
\end{itemize}
\noindent
All the three dimensions above play an important role on the answer quality. As an example, we report below the statistical analysis results of the spatial context, social context, and temporal context.

\begin{table}[hbt]
\centering 
\caption{Statistics of varying answer quality at different locations.}
\begin{tabular}{|l|c c|}
\hline
\rowcolor[gray]{0.8} \textbf{Spatial context} & \textbf{High-quality answers}     
                   & \textbf{Low-quality answers }           \\ \hline
Home Apartment/room           & 45.30\%  & 55.70\%  \\ 
Home Relatives                & 41.47\%   & 58.53\%  \\ 
House Friends/others          & 29.76\%  & 70.24\%  \\ 
University Classroom/library  & 52.02\%  & 47.98\%  \\ 
University Canteen & 40.00\%  & 60.00\%  \\ 
Restaurant/pub                & 28.87\%        & 71.13\%  \\ 
In the street & 39.58\%  & 60.42\%  \\ 
Another indoor place          & 26.93\%   & 73.07\%  \\ 
Another outdoor place         & 28.68\%    & 71.32\%   \\ \hline
\end{tabular}
\label{location}
\end{table}
\noindent
For instance, as shown in Table \ref{location}, high-quality answers constitute a minor fraction of 28.87\% in restaurants or pubs. In contrast, in academic settings such as classrooms or university libraries, high-quality answers constitute 52.02\%. Focusing on low-quality answers, restaurants have the highest percentage of low-quality answers at 71.13\%, likely due to the relaxed and enjoyable atmosphere, the participant would not pay attention to the questions, which led to incorrect answers. Friends’ houses rank third with a low-quality answer percentage of 70.24\%, as visiting friends or attending social gatherings often leads to positive interactions and experiences, inducing not to focus on the smartphone questions.
\begin{table}[htp]
\centering
\caption{Statistics of varying answer quality on the different days of the week.}
\begin{tabular}{|l|c c|}
\hline
\rowcolor[gray]{0.8} \textbf{Weekday} & \textbf{High-quality answers}     
                     & \textbf{Low-quality answers }           \\ \hline
Monday    & 46.94\% & 53.06\%                        \\  
Tuesday   & 43.83\%    & 56.17\%                        \\  
Wednesday & 43.19\%    & 56.81\%                        \\  
Thursday  & 40.52\%   & 59.48\%                        \\ 
Friday    & 39.31\%   & 60.69\%                        \\  
Saturday  & 44.46\%    & 55.54\%  \\  
Sunday    & 42.02\%     & 57.98\%  \\ \hline
\end{tabular}
\label{Week}
\end{table}
If we move to the temporal context and analyze the data collected during the 28 days of the experiment, we can distinguish various distinct patterns across different weekdays, as shown in Table \ref{Week}. Specifically, high-quality answers are most prevalent on Monday (46.94\%) and Saturday (44.46\%), with the lowest incidence observed on Friday (39.31\%). Conversely, the incidence of low-quality answers peaks on Friday (60.69\%) and Thursday (59.48\%). These findings suggest a correlation between the day of the week and answer quality states. Notably, the onset of the workweek and weekend (Monday and Saturday) witnesses a relative surge in high-quality answers. Similar considerations can be made for the social context, see Table \ref{withw}.

\begin{table}[htp]
\centering
\caption{Statistics of varying answer quality with different interact people.}
\begin{tabular}{|l|c c|}
\hline
\rowcolor[gray]{0.8} \textbf{social context} & \textbf{High-quality answers}     
                     & \textbf{Low-quality answers}           \\ \hline
Alone    & 45.66\% & 54.34\%                        \\  
Partner   & 35.50\%    & 64.50\%                        \\  
Roommates & 43.06\%    & 56.94\%                        \\  
Classmates  & 37.82\%   & 62.18\%                        \\ 
Relatives    & 45.21\%   & 54.79\%                        \\  
Friends  & 28.90\%    & 71.10\%  \\  
Colleagues/other    & 29.42\%     & 70.58\%  \\ \hline
\end{tabular}
\label{withw}
\end{table}

\begin{figure}[htp!]
\centering
\includegraphics[width=0.7\textwidth,height=0.3\textwidth]{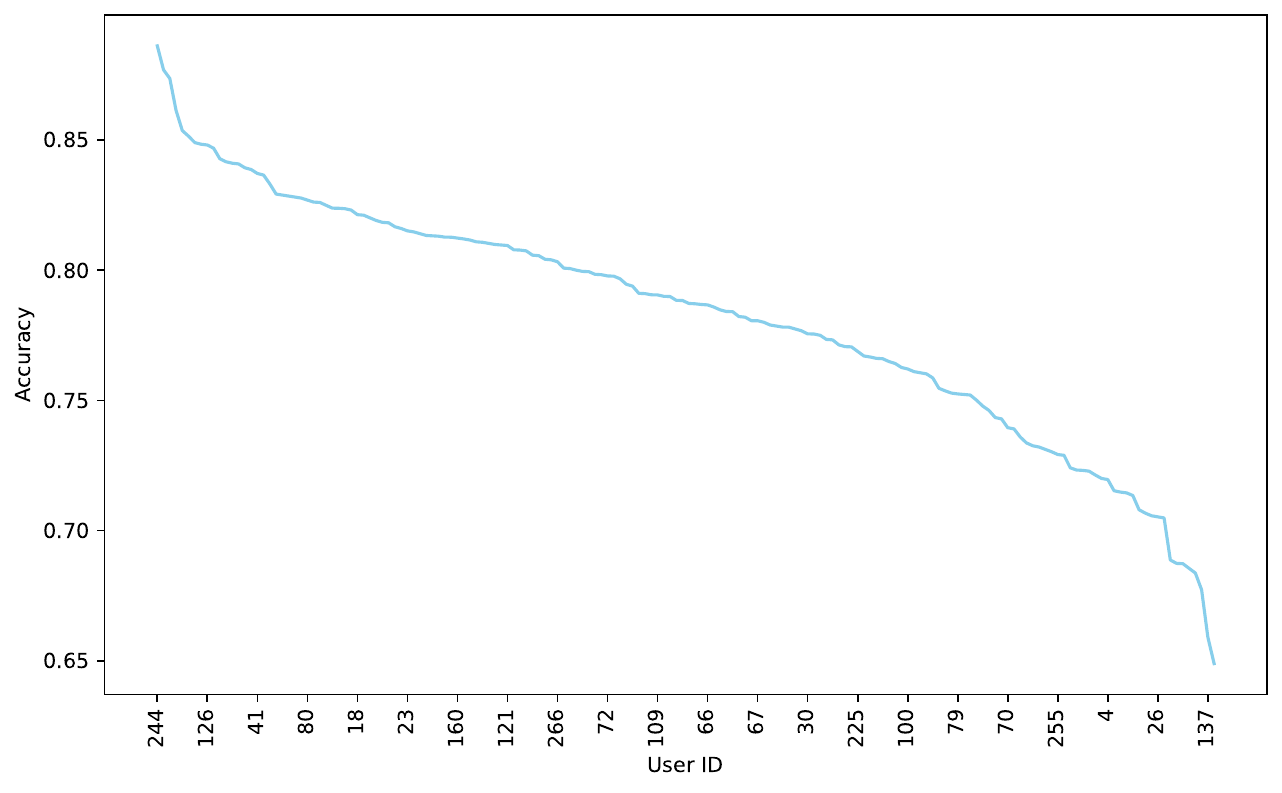}
\caption{ \centering The prediction results of each participant, with some participant ids made explicit.}
\label{person}
\end{figure}

\begin{table}
    \centering
    \caption{Prediction answer quality results of different machine learning classifiers.}
    \begin{tabular}{ |c|c c c c c| }
    \hline
     \rowcolor[gray]{0.4} \textbf{Classifier} &\textbf{Accuracy} &\textbf{Kappa} &\textbf{Precision} & \textbf{Recall} & \textbf{F1 score}\\
    \hline
     Random Forest&0.7583&0.4472&0.7092&0.7283&0.7200\\

     KNN&0.7311&0.4389&0.7023&0.7254&0.7229\\

     Logistic Regression&0.7017&0.3914&0.6595&0.7003&0.7002\\

     Gaussian Naive Bayes&0.6687&0.3405&0.6081&0.6687&0.6761\\

    SVM&0.7282&0.4097&0.6712&0.7082&0.7104\\
    \hline
    \end{tabular}
    \label{class}
\end{table}
\noindent
As part of the flexibility that the platform provides, we can apply various ML models on the data collected. As from Table \ref{class}, in this work we have used Random Forests (RF), K Nearest Neighbors (KNN), Logistic Regression, Support Vector Machines (SVM), and Gaussian Naive Bayes. The classifiers were trained using 5-fold cross-validation on the comprehensive training and testing sets for all participants, with 80\% of the data allocated to the training set and the remaining 20\% to the testing set. The goal was to predict whether the participant would answer the question within 30 minutes, this being the time within which the answer is most likely to be correct, as from \cite{bison2024impacts}.
As depicted in Table \ref{class}, Random Forest surpassed the other four classifiers, achieving a prediction accuracy of 0.758. These machine learning results demonstrate that our component can effectively predict if the participant can answer questions within 30 minutes, thus predicting answer quality based on the context information collected through questions and sensors. The Random Forest classifier has been selected for the next step, which was to predict the answer quality for each participant.
Focusing on each specific participant, we used their first two weeks of data to train the random forest algorithm and then predicted their answer quality in the subsequent two weeks. As illustrated in Fig \ref{person}, the accuracy of answer quality predictions varies for each participant. In this figure, the x-axis represents the user ID, which distinguishes different users, and the y-axis indicates the accuracy of predicting the moments when each user can answer questions with high quality. The figure is arranged in descending order of accuracy, meaning the first user achieves the best performance, and the last one faces the worst.
It is worthwhile to analyze the results in the case of a few selected participants as examples. We have the following:

\begin{itemize}
    \item Participant 244 demonstrated a high prediction accuracy of 88.61\%. This participant consistently provided high-quality answers during the periods identified by the algorithm. This participant was most of the time alone and at the university.
    \item Participant 30 showed a moderate prediction accuracy of 77.55\%. The variability in the context information contributed to the fluctuations in answer quality. This participant always provided a lot of different context information in different places and activities, which made it hard to predict.
    \item Participant 137: Exhibited the lowest prediction accuracy of 65.89\%. This participant provided a non negligible number of wrong answers, e.g., driving while being in the University classroom or in the library.
\end{itemize}
The overall conclusion is that we have a good average level of predictability which does not decay much in the worst case, with very good results in the best situations.

%% file: section/8-Discussion.tex
\section{Conclusion}
\label{discussion}

% In conclusion, 
This paper introduces a novel methodology and platform, an enhanced version of the iLog APP, for the collection of large-scale sensor data and qualitative human feedback. 
Our main contributions are as follows:

\begin{itemize}
    \item a language for modeling the situational context based on five key dimensions, that is: spatial, temporal, internal, social, and utensil;
    \item a language for modeling the temporal context, thus enabling a precise scheduling, still very flexible and modifiable during experiment execution, of the various aspects of the data collection;
    \item a dashboard component which enables both researchers and participants to edit and monitor the progress of the experiment plan as a prerequisite for enhancing the quality of the data collection;
    \item a machine learning component which allows the platform to infer about the best moment to take action, for instance, to ask a question. 
\end{itemize}
We foresee two avenues for future research. The first is the exploration of  ways to validate and verify the user responses and reduce the burden of answering questions.  To address this issue, the starting point will be the work on Skeptical Learning \cite{zhang2022skeptical,zhang2019personal} which allows handling of mislabeling in personal context recognition.  
    The second is the validation of the methodology of the platform in other domains with a specific interest in health and   entertainment.